\documentclass[]{pasj00}

\usepackage{lscape}
\usepackage{natbib}

\begin{document}

\title{An emergence of new polarized emission region in blazar Mrk~421 associated with X-ray flare}

\author{

Ryosuke \textsc{Itoh}\altaffilmark{1,2}, 
Yasushi \textsc{Fukazawa}\altaffilmark{1}, 
Yasuyuki~T. \textsc{Tanaka}\altaffilmark{3}, 
Koji~S. \textsc{Kawabata}\altaffilmark{3}, 
Katsutoshi \textsc{Takaki}\altaffilmark{1}, 
Kazuma Hayashi\altaffilmark{1},
Makoto \textsc{Uemura}\altaffilmark{3}, 
Takahiro \textsc{Ui}\altaffilmark{1}, 
Mahito \textsc{Sasada}\altaffilmark{4}
Masayuki \textsc{Yamanaka}\altaffilmark{5}  
and Michitoshi \textsc{Yoshida}\altaffilmark{3}
}

\altaffiltext{1}{Department of Physical Science, Hiroshima University, Higashi-Hiroshima, Hiroshima 739-8526, Japan}
\altaffiltext{2}{email: itoh@hep01.hepl.hiroshima-u.ac.jp}
\altaffiltext{3}{Hiroshima Astrophysical Science Center, Hiroshima University, Higashi-Hiroshima, Hiroshima 739-8526, Japan}
\altaffiltext{4}{Kwasan Observatory, Kyoto University, Ohmine-cho Kita Kazan, Yamashina-ku, Kyoto 607-8471, Japan}
\altaffiltext{5}{Department of Physics, Faculty of Science and Engineering, Konan University, Okamoto, Kobe, Hyogo 658-8501, Japan}
\maketitle

\begin{abstract}
We report on long-term multi-wavelength monitoring of blazar Mrk~421 from 2010 to 2011.
The source exhibited extreme X-ray flares in 2010. 
Our research group performed optical photopolarimetric follow-up observations 
using the Kanata telescope.  
In 2010, the variability in the X-ray band was significant, while the optical
and ultraviolet (UV) flux decreased gradually.
Polarization properties also exhibited unique variability in 2010,
suggesting the presence of systematic component of polarization 
and magnetic field alignment for the emergence of a new polarized 
emission region.
In contrast, in 2011 the variability in the X-ray band was smaller, and
the variability in the optical and UV bands was larger, than in 
2010.
To explore the reasons for these differences, spectral fitting analysis was performed via simple 
synchrotron-self Compton modelling; the results revealed different
behaviors in terms of spectral evolution between these periods, suggesting 
different variability mechanisms between 2010 and 2011. 
In 2010, the radiation was likely the result of energy injection into the emitting regions with an aligned magnetic field. 
In contrast, in 2011 the superposition of different emission regions 
may have contributed to the low degree of observed polarization.
It also implies that high-energy electron which were not accelerated to 
ultra-relativistic velocities were injected in 2011.  
\end{abstract}

\section{Introduction}
Blazars are highly variable active galactic nuclei (AGN) that can be detected for all
wavelengths, ranging from radio waves to gamma rays.  
The strong relativistic jets from blazars are aligned with the observer's line of 
sight and appear bright due to relativistic beaming, having 
a luminosity typically greater than $10^{43}$~erg~s$^{-1}$ \citep{1995PASP..107..803U}. 
Blazar emission typically consists of two spectral components: 
one is attributed to synchrotron radiation from relativistic electrons, emitted at lower energies peaking in the radio through optical regime; the 
second occurs at higher energies from inverse Compton scattering 
radiation, peaking in the gamma-ray regime.
A notable characteristic of blazars is their rapid, high-amplitude 
intensity variations or flares.
Blazars can be classified into three types, based on the peak synchrotron 
radiation frequency $\nu_{\rm peak}$ \citep{2011ApJ...743..171A}:
low-synchrotron-peaked blazars (LSP), intermediate-synchrotron-peaked blazars (ISP),
and high-synchrotron-peaked blazars (HSP).
Due to relativistic effects, the radiation from jets dominates the overall spectral energy distribution. 
As a result, the optical band spectra of blazars are essentially featureless, compared with other AGNs. This makes blazars highly suitable for jet study.

Polarized radiation from blazars, an indicator of low-energy synchrotron radiation, 
can vary sharply.
Thus, the polarization of blazars is of interest for understanding the origin, confinement, and propagation of jets, 
due to the dependence of the polarization on the jet's magnetic field structure
\citep[e.g.,][]{1998AJ....116.2119V}.
\cite{1990A&AS...83..183M} performed a large-sample study of blazars in the optical band 
and found that a high polarization degree (PD) and variability of polarization are common phenomena in blazars.  
\cite{2011PASJ...63..639I} reported statistical photopolarimetric observations of blazars with a daily timescale, 
and suggested that the lower luminosity and higher peak frequency of synchrotron radiation objects (e.g., an HSP blazar) have 
smaller amplitudes for variations in the flux, color, and polarization degree.
However, questions remain regarding the reason for the difference between HSP and LSP blazars.
What seems to be lacking is simultaneous multi-wavelength and polarimetric observations.
Multi-wavelength observation is valuable because the time-resolved spectral energy distribution (SED) in the order of physically relevant timescales 
allows verification of a synchrotron-self-Compton (SSC) model that considers the time-development of the electron spectra.
Wide-range, temporally deep, multi-wavelength polarization studies 
should reveal
some of the blazar's characteristics, such as its energy dependence, time variability, and jet structure.

Markarian~421 (Mrk~421; R.A. = $11^h 04^m 27.3^s$, Decl. = $+38\arcdeg\ 12\arcmin\ 31\arcsec.8$, J2000, z=0.031),
one of the most observed blazars, is classified as an HSP blazar \citep{2010ApJ...716...30A}.
Additionally, it was the first extragalactic source detected at TeV energies, over the range 0.5--1.5 TeV, by the 
Whipple telescopes \citep[e.g.,][]{1992Natur.358..477P}.
The SED exhibits a double-peaked feature with two maxima; its variations are 
basically 
well-understood within the framework of a simple one-zone SSC model \citep{1985ApJ...298..128B,2011ApJ...736..131A}.
Hence, Mrk~421 is a good target for flare mechanism studies about relativistic jets.
We monitored extreme activity in the 
Mrk~421 X-ray bands that occurred in 2010, via long-term photopolarimetric measurements with the Kanata telescope.
Here, we present our findings about the relationship between the variability observed in the synchrotron spectra and optical polarization measurements during the 2010 X-ray flare. 
We assumed a flat $\Lambda$CDM cosmology, with $H_{0}=75$ km s$^{-1}$Mpc$^{-1}$ and $\Omega_m=0.3$ \citep{2006PASP..118.1711W}.

\section{Observation}
\subsection{Observation using the Kanata telescope}

We performed {\it V} and {\it R$_C$}-band photometry and
polarimetry observations of Mrk~421
from December 2009 to May 2011, using the HOWPol instrument 
installed on the 1.5-m Kanata telescope located at the 
Higashi-Hiroshima Observatory, Japan \citep{2008SPIE.7014E.151K}.
We obtained 74 daily photometry data points for the {\it V} band,
and 68 photometry and polarimetry data points for the {\it R$_C$} band.
A sequence of photopolarimetric observations
consisted of successive exposures 
at four position angles of a half-wave plate: 
$0^{\circ}, 45^{\circ}, 22.5^{\circ}, $and $67.5^{\circ}$.
The data were reduced under standard charged coupled device (CCD) 
photometry procedures.
We performed aperture
photometry using \verb|APPHOT| packaged in \verb|PYRAF|
\footnote{PYRAF is a product of the Space Telescope Science Institute, 
which is operated by AURA for NASA http://www.stsci.edu/institute/software\_hardware/pyraf},
and differential photometry with a comparison star imaged in the
same frame as Mrk~421.
The comparison star is located at R.A. = 11:04:18.2 and Decl. = +38:16:30.5 (J2000);
its magnitudes are {\it V} = 15.57 and {\it R$_C$} = 15.20 \citep{1998A&AS..130..305V}.
The data were corrected
for Galactic extinctions $A(V)=0.050$ and $A(R)=0.040$.

Note that it is also important to correct the instrumental polarization caused by  
reflection at the tertiary mirror of the telescope. 
The instrumental polarization was modeled as a function of the declination of the object and the
hour angle of observation, which was subtracted from the observed value.
We confirmed that the accuracy of the instrumental polarization subtraction was smaller than 0.5\%
for the {\it R$_C$}  band, using unpolarized standard stars. 
The polarization angle (PA) was defined in the 
standard manner (measured from north to east), 
based on calibrations with polarized stars, HD183143 and HD204827 \citep{1983A&A...121..158S}.
Because the PA has an ambiguity of $\pm 180^{\circ} \times
n$ (where $n$ is an integer), 
we selected $n$, which gives the minimal angle 
difference from previous data, assuming that 
the PA changes smoothly.
We also confirmed that the systematic error caused by instrument 
polarization was smaller than $2^{\circ}$, using the polarized stars as mentioned.

\subsection{Observation using {\it Swift} UVOT and XRT}

{\it Swift} observed the source between 2009 December 15 and 2011 February 22.
A total of 125 data points in the {\it UVW2} band, 122 data points in the {\it UVM2} band, 
and 111 data points in the {\it UVW1} band were obtained with the UV and Optical Telescope (UVOT).
UVOT data were reduced following the standard procedure for CCD photometry.
Counts were extracted from an aperture of 5 arcsec radius for all
filters and converted to flux using the standard zero points.
\citep{2008MNRAS.383..627P}.
The data were corrected for Galactic extinctions 
$A(UVW2)=0.136$, $A(UVM2)=0.143$, and $A(UVW1)=0.098$ \citep{1998ApJ...500..525S}.

\begin{figure*}[!htb]
  \centering
  \includegraphics[angle=0,width=8cm]{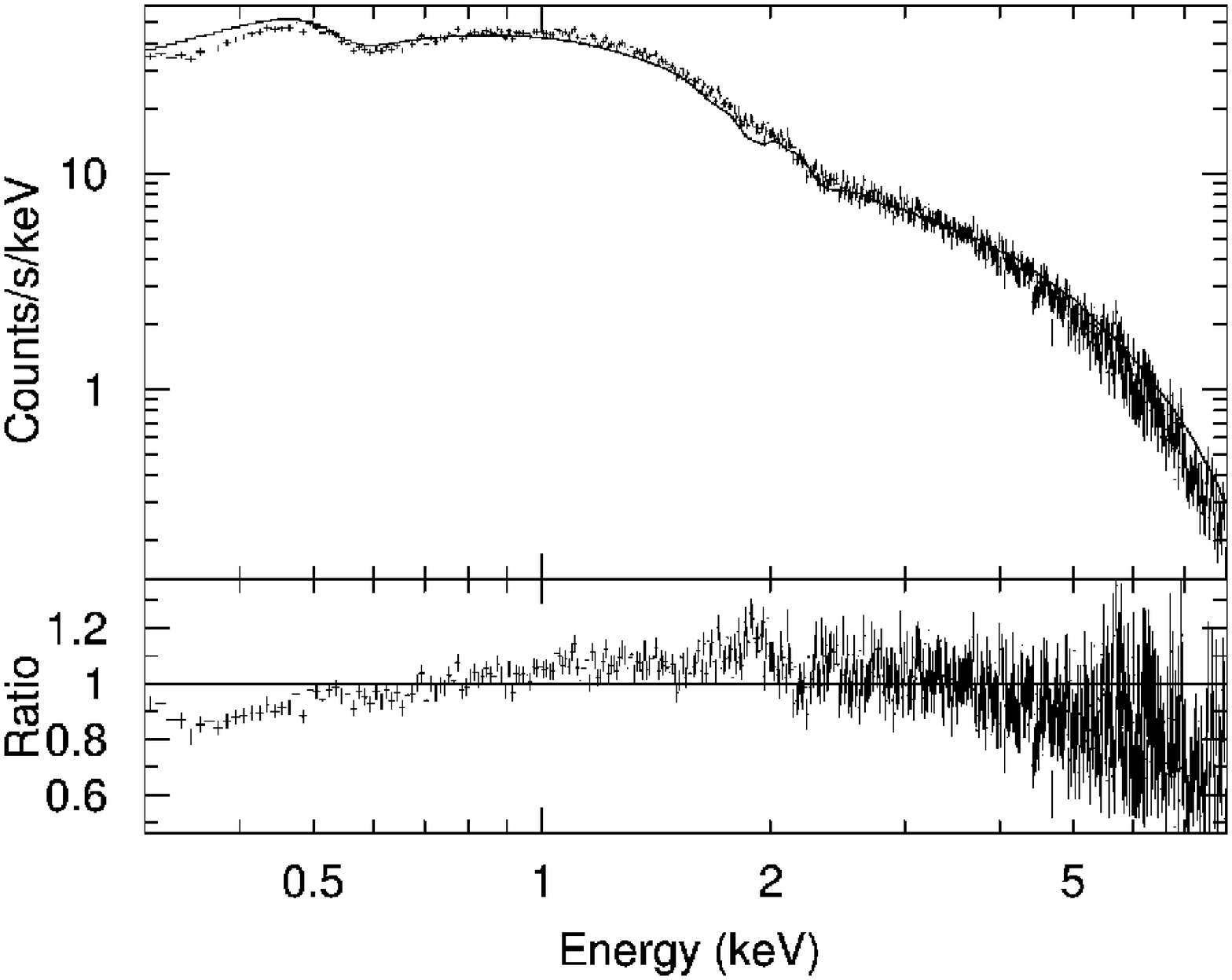}
  \includegraphics[angle=0,width=8cm]{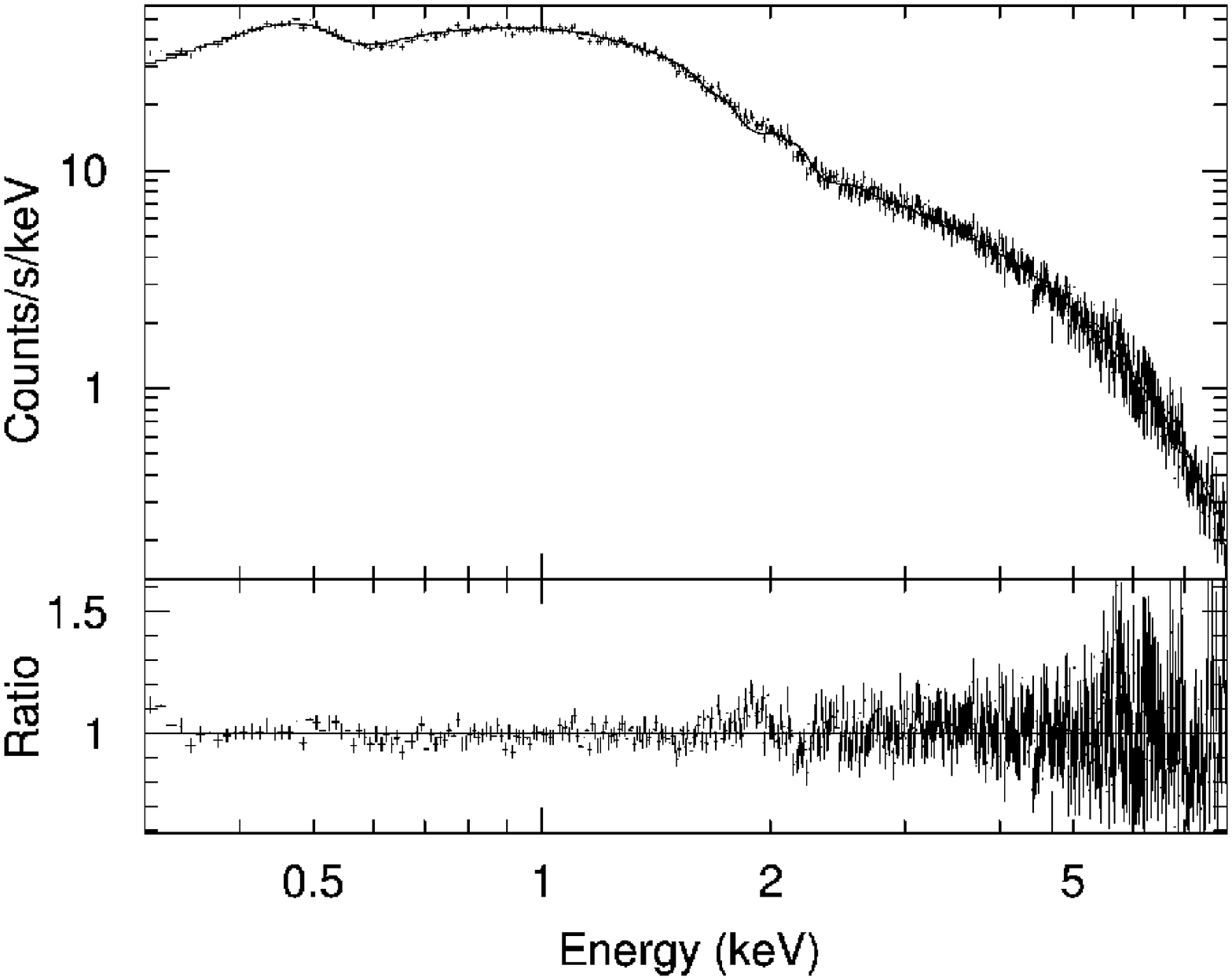}
  \caption{{\it Swift} X-ray Telescope (XRT) spectral analysis of Mrk~421 (MJD=55148). 
    Left: Spectral fit using the power-law model with systematic deviations on 
    both sides of the residuals.
    Right: Spectral fit with the log-parabolic model with no deviation.}
  \label{fig:xrtspec}
\end{figure*}

We also used data obtained by the X-Ray Telescope (XRT) in the same epoch.
There were 165 data points observed by XRT. 
All XRT data presented here were taken in windowed timing (WT) 
mode\footnote{see \cite{2005SSRv..120..165B} for more details on XRT observing modes}, 
due to the high flux rate of the source.
Data reduction and calibration of the XRT data were performed with HEASoft v6.4 
standard tools.
Events in the energy range from 0.2 -- 8 keV, with grades 0 -- 2, were used for the analysis. 
Source spectra were binned to ensure a minimum of 20 counts
per bin to facilitate $\chi^2$ minimization fitting.
Response files were generated with \verb|xrtmkarf|
Task, with corrections applied for point-spread function (PSF) losses and CCD defects.
Spectral analysis was performed with the \verb|XSPEC| software package version 12.
An estimate of the absorption at low energies, caused by interstellar gas, 
is required for spectral curvature analysis.
Because there was no evidence of the host galaxy in Mrk~421 by optical observation \citep{2000ApJ...532..816U},
we used an absorption model having a hydrogen-equivalent column density (N$_H$) similar to that of our galaxy, 
$N_H = 1.61 \times 10^{20}$cm$^{-2}$  \citep{1995ApJS...97....1L}.
We performed spectral analysis by fixing the N$_H$ absorbing column densities to the galactic values.
In \cite{2004A&A...413..489M} and \cite{2007A&A...466..521T}, the spectral shape of Mrk~421 in the X-ray band was 
well fitted with a log-parabolic model, as opposed to a power-law form.
Spectral fitting with the log-parabolic shape is described below:
\begin{equation}
F(E) = K E^{-a+b \log(E)} \ \   \textrm{ph cm}^{-2}\ \textrm{s}^{-1}\ \textrm{keV}^{-1}
\end{equation}
where $a$ is the photon index at 1 keV and $b$ is the spectral curvature.
Figure \ref{fig:xrtspec} shows the spectral fitting results for the power-law and log-parabolic models.
A $\chi^2_r$ test for the fit with a simple power-law model indicates a reduced value, $\chi^2_r$ = 3.57548 (d.o.f = 687), for the data taken on MJD=55148.
In contrast, the reduced $\chi^2_r$ value using the log-parabolic model showed that $\chi^2_r$ = 1.279 (d.o.f. = 686)
for the same data.
Improvement in the reduced $\chi^2_r$ values was also evident for other observational data. Therefore, we used the log-parabolic model in this paper.

\subsection{Observation using MAXI}

We used archival data\footnote{http://maxi.riken.jp/top/} of Mrk~421 using Monitor of All-sky X-ray Image/Gas Slit Camera (MAXI/GSC) observations.
GSC is designed to scan the entire sky every 92-min orbital period in the 2 -- 30 keV band, giving it 
the highest sensitivity of the all-sky X-ray monitors.
The data used in this analysis were taken between December 2009 and May 2011.
The energy range was set to 2 -- 20 keV in 1-day data bins.
Details of the data analysis of Mrk~421 for the 2010 January and February flares were
reported in \cite{2010PASJ...62L..55I}.

\section{Results}

\subsection{Light Curves}

\begin{figure*}[!htb]
  \centering
  \includegraphics[angle=0,width=15cm]{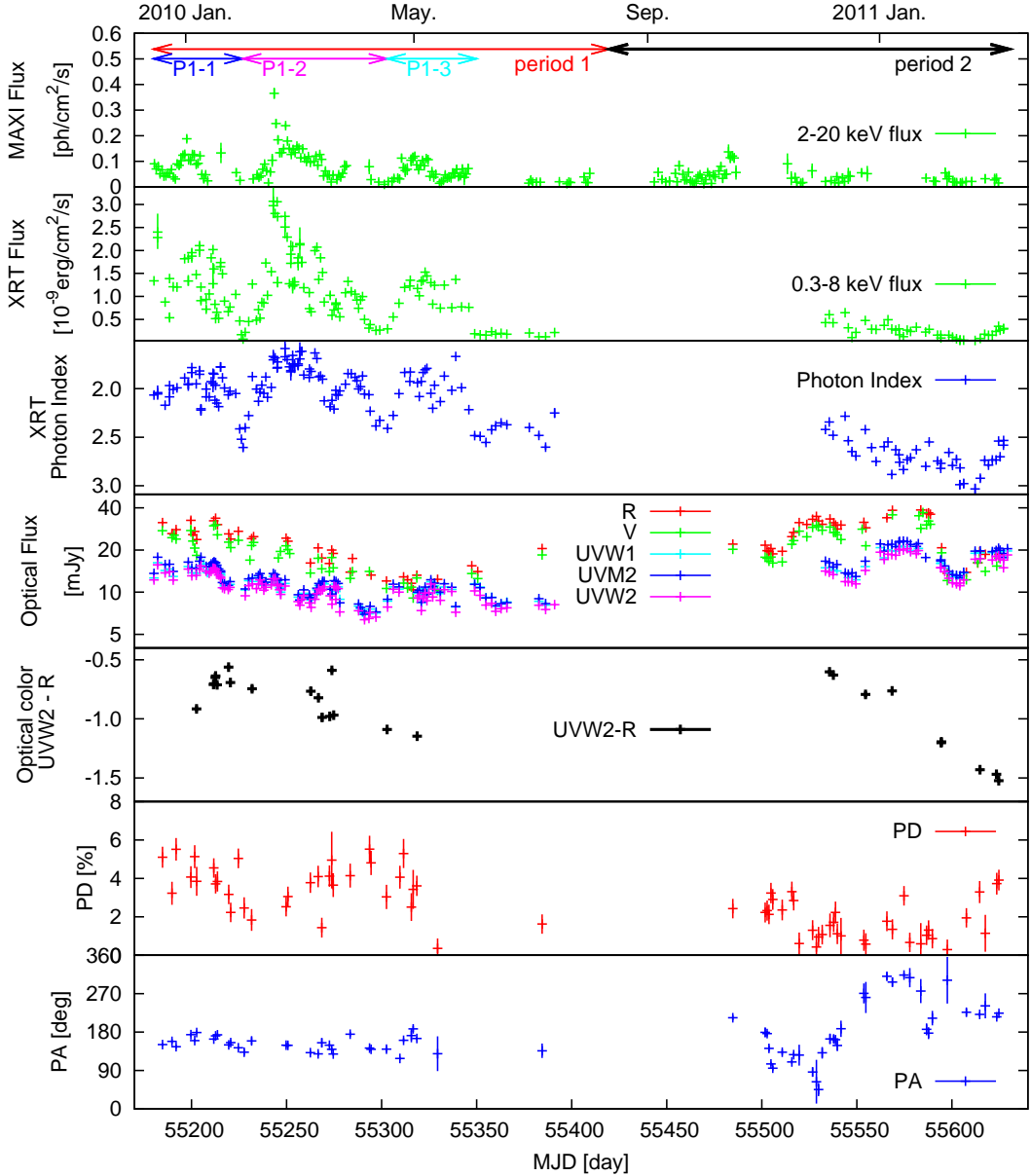}
  \caption{Light curves in multi-wavelength observation of Mrk~421 between Jan. 2010 and Feb 2011.
    The top panel shows 2 -- 20 keV flux measured by MAXI/GSC. 
    The second panel shows 0.3 -- 8 keV flux measured by {\it Swift}/XRT.
    The third panel shows the photon index at 1 keV, measured by {\it Swift}/XRT.
    The fourth panel shows the UV and optical flux measured by {\it Swift}/UVOT and Kanata/HOWPol.
    The fifth panel shows UV and optical color. 
    The sixth panel shows the polarization degree in the {\it R$_C$} band and the PA (bottom panel).
  }
  \label{fig:MWLC}
\end{figure*}

Figure \ref{fig:MWLC} shows the multi-wavelength light curves from the optical to X-ray bands between MJD 55180 and 55630 
(corresponding to 2009-12-15 and 2011-03-10). 
Below we describe details of the light curves for each energy band.

\subsubsection{X-ray Flux and Photon Index}
Photon flux at 2 -- 20 keV measured by MAXI/GSC and the energy flux at 0.3 -- 8 keV 
measured by {\it Swift}/XRT showed similar trends in variability.
In the first half of the observations ($< \textrm{MJD} \sim 55400$, hereafter referred to as 'period 1'), 
X-ray band variability was prominent compared with that in the last half ($> \textrm{MJD} \sim 55400$ ; 'period 2').
On MJD 55250, the X-ray flux reached $30.6^{+0.3}_{-0.3} \times 10^{-10}$ [erg cm$^{-2}$ s$^{-1}$] at its maximum value.
In period 2, however, the variability for the X-ray band was negligible.
The average flux in period 2 was $F_{2\textrm{;ave}} = 1.5 \times 10^{-10}$ [erg cm$^{-2}$ s$^{-1}$]. 
The maximum flux in period 1 and average X-ray flux values for period 2 differed by a factor of $\sim 20$.
MAXI observations indicated that there were no large flares from MJD 55400 to 55530; however, the observation of {\it Swift} 
was limited by the location of Mrk~421, close to the sun.
The photon index also showed clear variability in period 1, indicative of the trend of 'harder when brighter'.
A similar trend was reported in \cite{2009A&A...501..879T} for this source for previous X-ray flares.
Note that photon indices for period 2 were softer than the values for period 1.

\subsubsection{UV and Optical-band Flux and Color}
UV and optical fluxes also showed clear variability, but their trends were different from that exhibited by the X-ray flux.
In period 1, flux variability gradually decreased, with no significant flares correlated with the X-ray flare.
In period 2, the UV and optical flux changed significantly.
The {\it R$_C$}-band maximum and minimum flux differed by a 
factor of $\sim 3$ in period 2, while the 
X-ray flux remained nearly constant.
The UV and optical color were statistically compatible and relatively constant. 
The spectrum shape variability of the UV and optical bands was smaller than that for the X-ray band.

\subsubsection{Optical Polarization}
\begin{figure*}[!htb]
  \centering
  \includegraphics[angle=0,width=8cm]{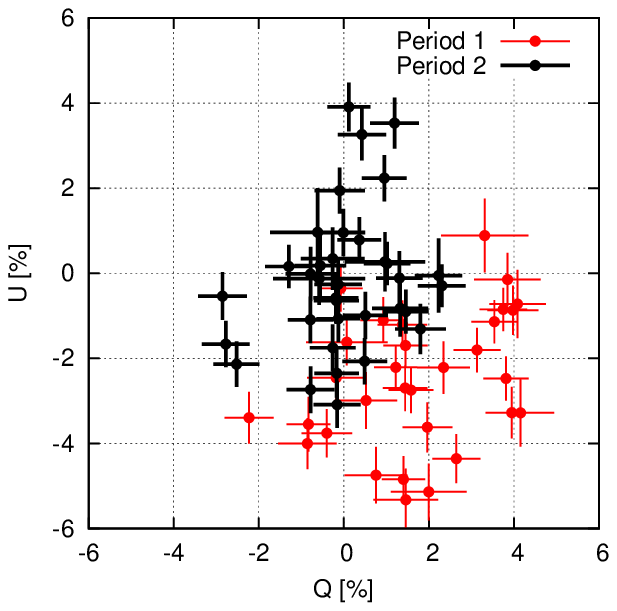}
  \includegraphics[angle=0,width=8cm]{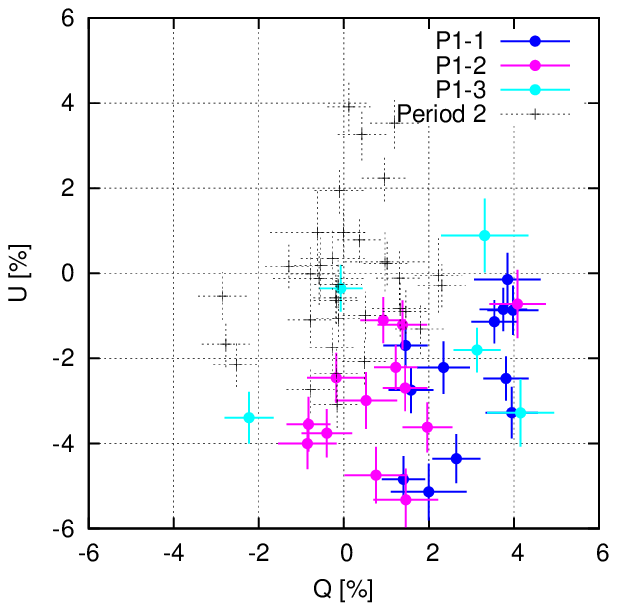}
  \caption{Polarization on Stokes QU plane for each period. 
    In the left figure, red indicates period 1 (MJD 55180--55430) and 
    black indicates period 2 (MJD 55431--55630).
    In the right figure, the data points in period 1 are separated into three epochs: 
    (P1-1, MJD 55180--55227; P1-2, MJD 55227--55303; and P1-3, MJD 55303--55350).
  }
  \label{fig:QU}
\end{figure*}

With regard to polarization, the PD exhibited large variability in period 1. 
Similar to the variability of the X-ray flux, 
the variability in PD for period 1 was larger than that observed for period 2.
In period 2, the PD was low and inactive. 
In contrast, the variability of the PA was small for period 1.
The average value of PA for period 1 was $\sim 150$ degree. 
As opposed to the variability of PA observed in period 1, 
The PA in period 2 changed significantly. 
Such a rotation of the PA was similar to that observed for BL~Lac and 3C~279
\citep{2008Natur.452..966M,2010Natur.463..919A}; however, this observation requires closer examination, due to the 
pseudo-rotation of PA that appears when the PD is low. 
Figure \ref{fig:QU} shows the polarization on the Stokes parameter QU plane for each period.
Variation on the QU plane implies rotation in period 2; however, this variation could also be attributed to random motion about the origin of the QU plane.

The variation on the QU plane suggests a difference in the polarization 
between periods 1 and 2.
In figure \ref{fig:QU}, the locations of Q and U were systematically different for periods 1 and 2:   
specifically, there was some systematic component of polarization in period 1.
The average values of Q and U for each period were $\textrm{Q}_1 = 0.013 \pm 0.005$,  $\textrm{U}_1 = -0.019 \pm 0.005$, 
$\textrm{Q}_2 = 0.000 \pm 0.005$, and $\textrm{U}_2 = -0.004 \pm 0.005$, respectively.
To examine the variability in more detail for period 1,
we separated the data points in period 1 into three epochs, based on the X-ray light curve (also see figure \ref{fig:MWLC}): 
P1-1 (MJD 55180--55227), P1-2 (MJD 55227--55303), and P1-3 (MJD 55303--55350).
The right panel of figure \ref{fig:QU} shows the detailed behavior of variability of QU in period 1.
These results indicate that the variability of PA in each flare has a large 
variance.
In other words, the PA did not appear to be completely constant during X-ray flares.
We also checked for the presence of a systematic component of polarization, as reported by \cite{2014AJ....148...42M}.
for period 2, but no significant component was confirmed.

\section{Discussion}

We performed photopolarimetric observations of Mrk~421 in an X-ray active state between 2010 and 2011.
Simultaneous wide-band spectra of synchrotron radiation were obtained on a daily timescale.
Generally, HSP blazars, such as Mrk~421, tend to not have good correlation between the polarization degree
and flux in the optical band \citep[e.g.,][]{2011PASJ...63..639I}.
The correlation between the X-ray flux and optical flux is also unclear for HSP blazars \citep[e.g.,][]{2008ApJ...686..181F}; sometimes they are correlated and sometimes not.
Therefore, our results suggest that we should take into account the type of flare when investigating corresponding changes in the X-ray band flux.

In the case that individual X-ray flares in period 1 were produced by a single emission region, 
a constant PA and good correlation between the PD and flux were expected.
Our observations, however, indicated some variation in the PA; additionally, no clear correlation was evident between the flux and PD using a timescale of days.
Systematic component of polarization was detected in period 1 (also see figure 
\ref{fig:QU}).
Taken together, these results suggest that the individual X-ray flares may be the result of the superposition of a long-term component (timescale of weeks to months) and several short-term components (timescale of days), as opposed
to a model using the superposition of short-term components without 
the long-term component \citep[e.g.,][]{1982ApJ...260..415M,2010PASJ...62...69U}.
This would explain the systematic difference of QU between period 1 and period 2, 
as well as the complexity of the correlation between flux and PD.

The emergence of a new polarized emission region associated with the long-term component accompanied by an X-ray flare in period 1 also indicates that 
such extreme X-ray flares originate in a region where the magnetic field is aligned.
Shocks in the jets lead to variations in the local emissivity and produce outbursts.
Due to the compression of the magnetic fields during shocks, these variations will also lead to
variations in the polarization.
Additionally, the average PA of ~150 degrees, without any large changes in PA such as the rotation or swing observed in
period 2, is in good agreement with the position angle of the
parsec-scale jet measured in the radio band \citep{2005ApJ...622..168P}.
The conformity of the position angle of the jet and the PA have been shown 
for several blazars, suggesting a 'shock-in-jet' model \citep[e.g.][]{2008ApJ...672...40H,2013ApJ...775L..26I}.
Namely, the magnetic field, compressed by shocks propagating down the jet, becomes aligned in a 
direction transverse to the jet.
Note that long-term component of PA also suggests the presence of a persistent magnetic field in the 
underlying jet, because the timescale of the long-term quasi-stationary PA is somewhat long. Small variations in the polarization may be explained by the superposition of different subcomponents.
Details of this scenario are discussed later in this section.

In contrast to the behavior in period 1, the X-ray flux and polarization activity 
were 
low in period 2, while the optical flux showed large variability.
This implies that the variability mechanism was different between periods 1 and 2.
Thus, 
we attempted to resolve the difference between these periods using broad-band spectral fitting.
The emission mechanism of X-ray flares is well studied by modeling the SED with the synchrotron and SSC model.
Here, we researched the evolution of the SED for periods 1 and 2 by spectral fitting with a simple SSC model. 
The simplest leptonic model typically used to describe the emission from an HSP blazar is the one-zone SSC model \cite[][]{2008ApJ...686..181F}.
As mentioned above, the SED in period 1 may be consistent with a two-component model.
However, for the sake of simplicity, we fitted the data using a one-zone model for each period and compared the jet parameters 
of periods 1 and 2.

In the model, radio to X-ray emission is produced by synchrotron 
radiation from electrons in a homogeneous, randomly oriented magnetic field ($B$),
and gamma-rays are produced by inverse Compton scattering.
We parametrized the electron distribution $N_e(\gamma)$ as a broken power law, expressed as 
\begin{equation}
N_e(\gamma) = K_e \times 
\left\{ \begin{array}{ll}
  \left( \frac{\gamma}{\gamma_b} \right)^{-p1} & \gamma < \gamma_b \\
  \left( \frac{\gamma}{\gamma_b} \right)^{-p2} & \textrm{otherwise}, 
  \end{array} \right.
\end{equation}
where $\gamma$ is the Lorentz factor of the electron, $\gamma_b$ is the break energy of the electron, and 
$K_e$ is a normalization term for the electron spectrum within the electron Lorentz
factor from $\gamma_{\textrm{min}}$ to $\gamma_{\textrm{max}}$.
Note that we also tried to fit the spectra using a single power-law model of the electron distribution, but the fitting results showed physically incorrect values (e.g., the Doppler factor was $> 100$ for the single power-law model).
Similar results were reported in \cite{2008ApJ...686..181F}.
Thus, we selected the broken power-law model for use in this study.
Because our investigation focused on the variability of synchrotron emission, 
the gamma-ray data is not that relevant, and has therefore be only partially considered.
Therefore, we fixed $B$ to 0.038 G, derived from SED fitting of 
TeV gamma-ray data in \cite{2011ApJ...736..131A}.
SED data were selected with the condition that XRT and optical/UV 
data were simultaneously taken within a day.
With this selection, we obtained 72 data points from our observations.
To account for inverse Compton scattering, 
we used GeV gamma-ray data observed by the {\it Fermi} Gamma-ray Space Telescope.
\cite{2011ApJ...736..131A} reported that the GeV gamma-ray variability is small, so we assumed that the gamma-ray flux was constant in our fitting.
In fact, GeV gamma-ray variability coincident with the radio band has been reported \citep{2014A&A...571A..54L}.
In this study, however, gamma-ray variability was not essential for the fitting process, because our fitting focused on the variability of 
synchrotron radiation. Therefore, we performed SED fitting under the assumption of a constant GeV flux.
We used the 2FGL catalog value from 2008--2010 observations for our fitting \citep{2012ApJS..199...31N}.

\begin{figure*}[!htb]
  \centering
  \includegraphics[angle=0,width=12cm]{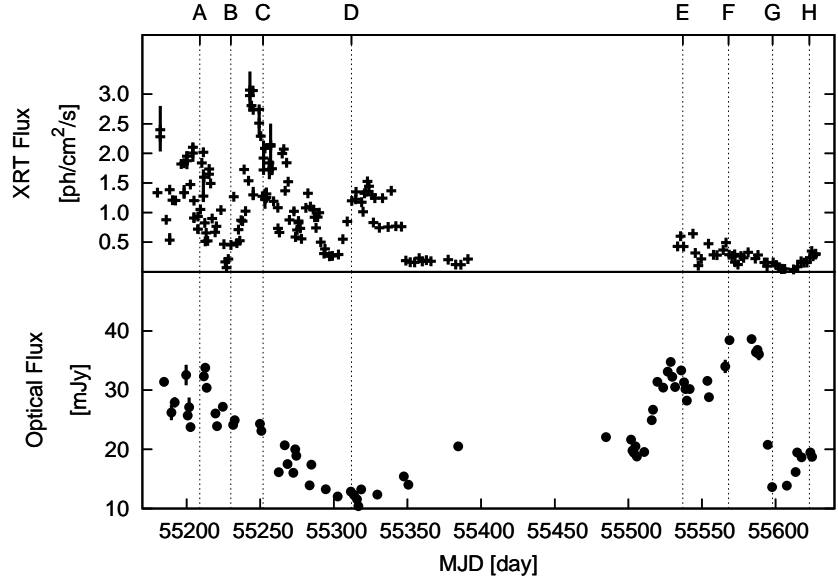}
  \caption{Light curves for X-ray and optical flux. The dotted line indicates dates 
    corresponding to SED fitting results shown in this paper.}
  \label{fig:LC_SED}
\end{figure*}

\begin{figure*}[!htb]
  \centering
  \includegraphics[angle=0,width=7cm]{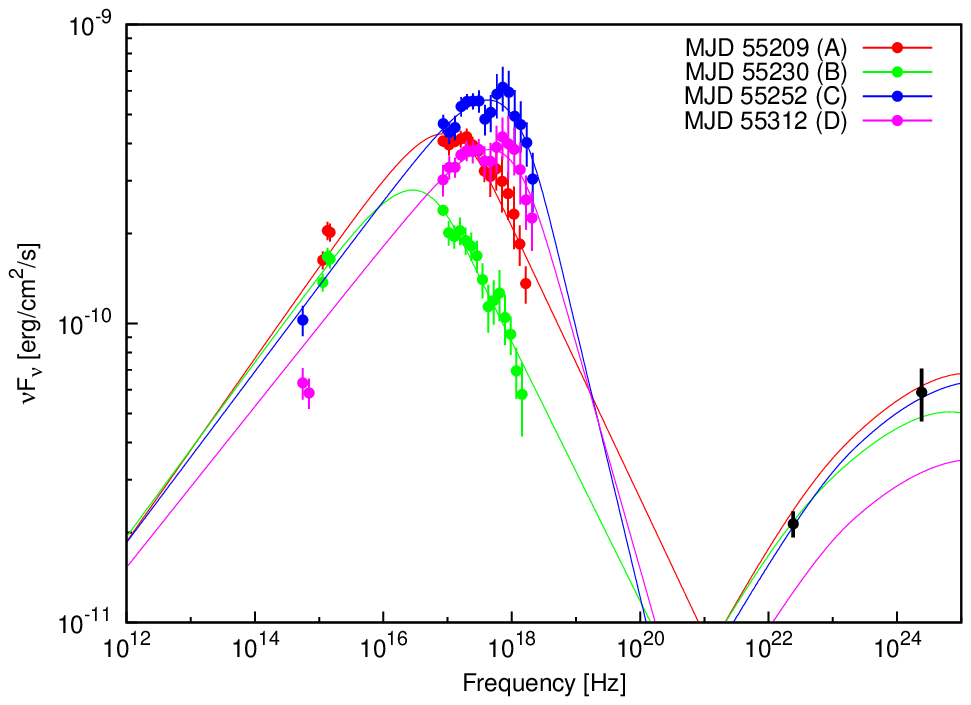}
  \includegraphics[angle=0,width=7cm]{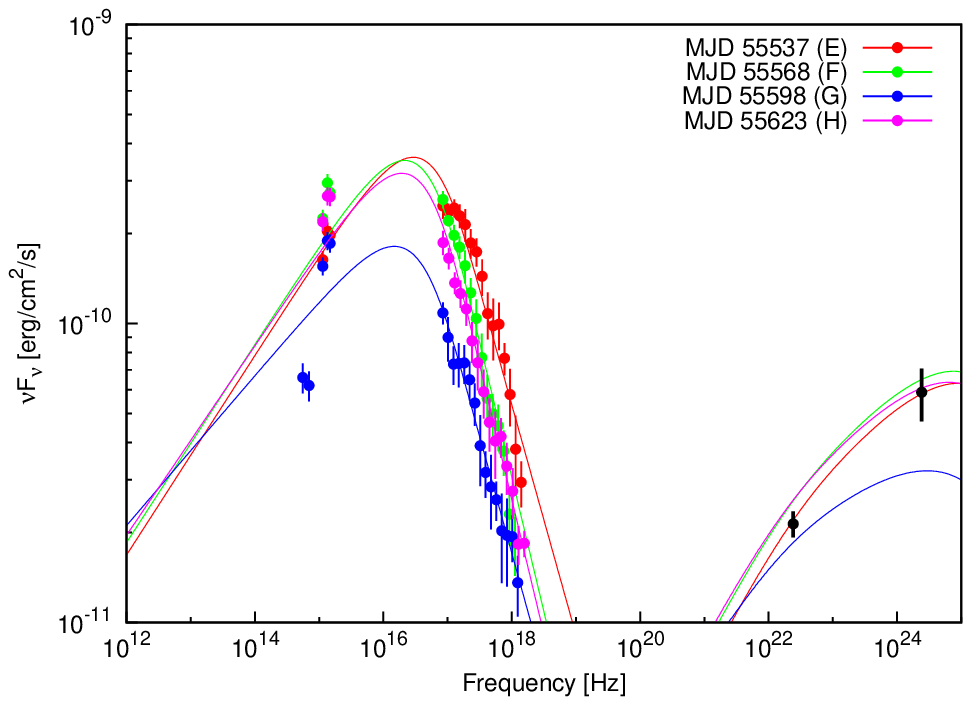}
  \caption{SEDs of Mrk~421 with SSC model fitting with the fixed parameters of $B$, $D$ and $R^{\prime}$ in period 1 and 2. 
    Colors indicate the four periods.
    The parameter values are shown in table \ref{tab:SED_BDT}.}
  \label{fig:SED_BDT_p1}
\end{figure*}

Figure~\ref{fig:LC_SED} shows the light curves for X-ray and optical data.
We performed SED fitting for different X-ray flux states, corresponding 
to the dates labeled in Fig~\ref{fig:LC_SED}. 
The one-zone SSC model provides a general description of the SED for both
data in periods 1 and 2.
However, upon closer examination, the one-zone SSC model may be insufficient for explaining simultaneously optical-UV and X-ray variability.
Specifically, these results suggest that the optical-UV and X-ray emission may be coming from different jet regions.
We first discussed the simplest one-zone SSC model. Next, we will
discuss the multi-emission model.
The parameter values derived by SED fitting are similar 
to the values reported in \cite{2011ApJ...736..131A}. The variability in the 
Doppler factor and the co-moving blob radius $R^{\prime}$ were small 
(e.g., the variability of the Doppler factor was $\sim \pm20\%$). 
To explore the variability in these periods in detail, we also performed SED 
fitting using a fixed parameter for the Doppler factor $D=21$ and co-moving blob radius 
$R^{\prime} = 5.2 \times 10^{16}$cm \citep{2011ApJ...736..131A}, because these parameters did not
show large variability in either of the two periods.
Hereafter, we described the results of SED fitting with fixed parameters for $B$, $D$, and $R^{\prime}$.
Figure~\ref{fig:SED_BDT_p1} and table~\ref{tab:SED_BDT}
show the SED fitting results in periods 1 and 2. 
As a result, the SEDs well described the synchrotron component.
Next, we discuss the variability mechanism of synchrotron emission 
based on these fitting results.
In period 1, a SED with a higher synchrotron-peak frequency tended to have a higher luminosity.
In contrast, in period 2, a SED with a lower synchrotron-peak frequency 
tended to have a higher luminosity.
These results suggest that the variability mechanisms were different for periods 1 and 2.

\begin{table}[!htb]
  \centering
  \caption{Fitting parameters with the fixed parameters $B$, $D$, and $R^{\prime}$ in period 1 and period 2 (Figure \ref{fig:SED_BDT_p1}).}
  \label{tab:SED_BDT}
  \begin{tabular}{ccccc}\hline\hline
    Date          & $p1^a$ & $p2^b$ & $Ke^c$                 & $\gamma_b^d$          \\ \hline
    MJD 55209 (A) & 2.39   & 3.91   & $103.4 \times 10^{40}$ & $19.0 \times 10^{4}$  \\
    MJD 55230 (B) & 2.42   & 3.87   & $350.0 \times 10^{40}$ & $10.9 \times 10^{4}$  \\
    MJD 55252 (C) & 2.43   & 4.80   & $7.5   \times 10^{40}$ & $50.8 \times 10^{4}$  \\
    MJD 55312 (D) & 2.46   & 4.51   & $5.0   \times 10^{40}$ & $50.7 \times 10^{4}$  \\ \hline
    MJD 55537 (E) & 2.33   & 4.54   & $352.6 \times 10^{40}$ & $12.2 \times 10^{4}$  \\
    MJD 55568 (F) & 2.34   & 4.80   & $500.0 \times 10^{40}$ & $10.8 \times 10^{4}$  \\
    MJD 55598 (G) & 2.50   & 4.56   & $354.7 \times 10^{40}$ & $9.5  \times 10^{4}$  \\
    MJD 55623 (H) & 2.37   & 4.80   & $500.0 \times 10^{40}$ & $10.4 \times 10^{4}$  \\
    \hline
  \end{tabular}
  \\
  {\footnotesize a: Index of the electron distribution ($ \gamma < \gamma_b$), b: Index of the electron distribution ($ \gamma > \gamma_b$),
  c: Normalization of the electron distribution, d: Break energy of the electron}
\end{table}

The variability of the synchrotron peak frequency to higher energy in period 1 
was well described by the scenario of high-energy electron injection into the 
blob. 
Fitting results of the break energy of the electron distribution also indicate 
high-energy electron injection in period 1.
A scenario with energy injection by internal shocks in relativistic shells (shock-in-jet model) is one way to explain such variability on a daily or monthly basis 
\cite[e.g.,][]{2001MNRAS.325.1559S,2002ApJ...572..762Z}.
The high peak in the synchrotron radiation could be obtained by the injection of energy into the emitting regions,
and a consequent increase in the number of high-energy electrons.
Polarimetric observation supports this scenario.
The relatively high polarized emission in period 1 is well represented by the
shock-in-jet model, which naturally results in a high PD, due to blob compression.
The shock-in-jet model also predicts the polarization vector orientation parallel with respect to the projected jet with compressed shocks.
As mentioned previously, the position angle of the parsec-scale jet was measured to be $-20$ deg, 
a value close to the average PA of 150 deg in period 1.
The reason for the small variability of optical flux in period 1 compared with that of X-rays was that 
the optical emission comes from electrons with less energy than $\gamma_b$.
Decreasing the normalization value of the electron distribution suggests that the variability of the long-term component 
could be initiated by decreasing the number of low-energy electrons.

The spectral variability in period 2 showed quite different behavior from that in period 1.
The variability in the X-ray band was relatively weak, compared with that of period 1.
From the right panel of Figure~\ref{fig:SED_BDT_p1} and Table~\ref{tab:SED_BDT}, 
we were able to estimate the variability mechanism in period 2.
One remarkable feature of the variability in period 2 was 
the increase in the normalization of the electron distribution.
This finding suggests that only the number of low-energy electrons increased, while changes in the spectral shape of the electron distribution were small in period 2. 
To reproduce such variability, we proposed a scenario in which 
insufficient acceleration or electron diffusion had a greater influence in period 2, compared with that in period 1.
Superposition of the multi-emission region indicated weak polarized emission in period 2.
In this case, electron diffusion was facilitated by weak confinement.
\cite{2010ApJ...724.1509U} discussed the possibility of 
a diffusive acceleration of electrons with subluminal shocks.
A turbulent multi-emission zone model provides another possible explanation \citep[e.g.,][]{2014ApJ...780...87M}.
As mentioned above, SED fitting with the one-zone model is limited in its ability to simultaneously present
optical-UV and X-ray variability.
The multi-emission model also explains the small variations in polarization in period 1 \citep[also see][]{2014A&A...571A..54L}.
In this case, the superposition of turbulent plasma downstream of the 
jet may explain the random behavior of the polarization.
This model also explains the increase in the number of low-energy electrons, under the assumption of an increase in the diffuse emission region downstream of the jet.

The SED model can also be constructed with a weak magnetic field (e.g., $B < 0.038$) instead of assuming a strong beaming factor.
However, we cannot verify which model is more suitable, due to the coupling parameter. 
To distinguish between these models, temporally dense monitoring of this source is required
for spectral evolutions predicted using a timescale of a few hours.
Thus, temporally and energetically dense monitoring of this source would enable a more detailed 
study, using two-component models and time-dependent SSC scenarios.

\section{Summary and Conclusion}

To summarize our multi-wavelength observations of Mrk~421, 
systematic differences in the polarization 
and variability mechanisms were observed between periods 1 and 2.
The systematic component of polarization indicated the emergence of a
long-term component in period 1 with an aligned magnetic field. 
The variability of the whole-band spectra in period 1 was explained by the
scenario of high-energy electron injection, such as that proposed by the
shock-in-jet model, which also reproduced the polarized emission
and direction of the PA. The variability observed in period 2 can also be 
described by high-energy electron injection; however, in this case, the electrons were not accelerated to ultra-relativistic velocities.  

\section{Acknowledgments}

This work was supported by the Japan Society for the Promotion of Science (JSPS).


\begin{thebibliography}{37}
\expandafter\ifx\csname natexlab\endcsname\relax\def\natexlab#1{#1}\fi

\bibitem[{{Abdo} {et~al.}(2010{\natexlab{a}}){Abdo}, {Ackermann}, {Ajello},
  {Axelsson}, {Baldini}, {Ballet}, {Barbiellini}, {Bastieri}, {Baughman},
  {Bechtol}, \& et~al.}]{2010Natur.463..919A}
{Abdo}, A.~A., {et~al.} 2010{\natexlab{a}}, \nat, 463, 919

\bibitem[{{Abdo} {et~al.}(2010{\natexlab{b}}){Abdo}, {Ackermann}, {Agudo},
  {Ajello}, {Aller}, {Aller}, {Angelakis}, {Arkharov}, {Axelsson}, {Bach}, \&
  et~al.}]{2010ApJ...716...30A}
---. 2010{\natexlab{b}}, \apj, 716, 30

\bibitem[{{Abdo} {et~al.}(2011){Abdo}, {Ackermann}, {Ajello}, {Baldini},
  {Ballet}, {Barbiellini}, {Bastieri}, {Bechtol}, {Bellazzini}, {Berenji}, \&
  et~al.}]{2011ApJ...736..131A}
---. 2011, \apj, 736, 131

\bibitem[{{Ackermann} {et~al.}(2011){Ackermann}, {Ajello}, {Allafort},
  {Antolini}, {Atwood}, {Axelsson}, {Baldini}, {Ballet}, {Barbiellini},
  {Bastieri}, {Bechtol}, {Bellazzini}, {Berenji}, {Blandford}, {Bloom},
  {Bonamente}, {Borgland}, {Bottacini}, {Bouvier}, {Bregeon}, {Brigida},
  {Bruel}, {Buehler}, {Burnett}, {Buson}, {Caliandro}, {Cameron}, {Caraveo},
  {Casandjian}, {Cavazzuti}, {Cecchi}, {Charles}, {Cheung}, {Chiang},
  {Ciprini}, {Claus}, {Cohen-Tanugi}, {Conrad}, {Costamante}, {Cutini}, {de
  Angelis}, {de Palma}, {Dermer}, {Digel}, {Silva}, {Drell}, {Dubois},
  {Escande}, {Favuzzi}, {Fegan}, {Ferrara}, {Finke}, {Focke}, {Fortin},
  {Frailis}, {Fukazawa}, {Funk}, {Fusco}, {Gargano}, {Gasparrini}, {Gehrels},
  {Germani}, {Giebels}, {Giglietto}, {Giommi}, {Giordano}, {Giroletti},
  {Glanzman}, {Godfrey}, {Grenier}, {Grove}, {Guiriec}, {Gustafsson},
  {Hadasch}, {Hayashida}, {Hays}, {Healey}, {Horan}, {Hou}, {Hughes},
  {Iafrate}, {J{\'o}hannesson}, {Johnson}, {Johnson}, {Kamae}, {Katagiri},
  {Kataoka}, {Kn{\"o}dlseder}, {Kuss}, {Lande}, {Larsson}, {Latronico},
  {Longo}, {Loparco}, {Lott}, {Lovellette}, {Lubrano}, {Madejski}, {Mazziotta},
  {McConville}, {McEnery}, {Michelson}, {Mitthumsiri}, {Mizuno}, {Moiseev},
  {Monte}, {Monzani}, {Moretti}, {Morselli}, {Moskalenko}, {Murgia},
  {Nakamori}, {Naumann-Godo}, {Nolan}, {Norris}, {Nuss}, {Ohno}, {Ohsugi},
  {Okumura}, {Omodei}, {Orienti}, {Orlando}, {Ormes}, {Ozaki}, {Paneque},
  {Parent}, {Pesce-Rollins}, {Pierbattista}, {Piranomonte}, {Piron}, {Pivato},
  {Porter}, {Rain{\`o}}, {Rando}, {Razzano}, {Razzaque}, {Reimer}, {Reimer},
  {Ritz}, {Rochester}, {Romani}, {Roth}, {Sanchez}, {Sbarra}, {Scargle},
  {Schalk}, {Sgr{\`o}}, {Shaw}, {Siskind}, {Spandre}, {Spinelli}, {Strong},
  {Suson}, {Tajima}, {Takahashi}, {Takahashi}, {Tanaka}, {Thayer}, {Thayer},
  {Thompson}, {Tibaldo}, {Tinivella}, {Torres}, {Tosti}, {Troja}, {Uchiyama},
  {Vandenbroucke}, {Vasileiou}, {Vianello}, {Vitale}, {Waite}, {Wallace},
  {Wang}, {Winer}, {Wood}, {Wood}, \& {Zimmer}}]{2011ApJ...743..171A}
{Ackermann}, M., {et~al.} 2011, \apj, 743, 171

\bibitem[{{Band} \& {Grindlay}(1985)}]{1985ApJ...298..128B}
{Band}, D.~L., \& {Grindlay}, J.~E. 1985, \apj, 298, 128

\bibitem[{{Burrows} {et~al.}(2005){Burrows}, {Hill}, {Nousek}, {Kennea},
  {Wells}, {Osborne}, {Abbey}, {Beardmore}, {Mukerjee}, {Short}, {Chincarini},
  {Campana}, {Citterio}, {Moretti}, {Pagani}, {Tagliaferri}, {Giommi},
  {Capalbi}, {Tamburelli}, {Angelini}, {Cusumano}, {Br{\"a}uninger}, {Burkert},
  \& {Hartner}}]{2005SSRv..120..165B}
{Burrows}, D.~N., {et~al.} 2005, \ssr, 120, 165

\bibitem[{{Finke} {et~al.}(2008){Finke}, {Dermer}, \&
  {B{\"o}ttcher}}]{2008ApJ...686..181F}
{Finke}, J.~D., {Dermer}, C.~D., \& {B{\"o}ttcher}, M. 2008, \apj, 686, 181

\bibitem[{{Hagen-Thorn} {et~al.}(2008){Hagen-Thorn}, {Larionov}, {Jorstad},
  {Arkharov}, {Hagen-Thorn}, {Efimova}, \& {Marscher}}]{2008ApJ...672...40H}
{Hagen-Thorn}, V.~A., {Larionov}, V.~M., {Jorstad}, S.~G., {Arkharov}, A.~A.,
  {Hagen-Thorn}, E.~I., {Efimova}, N.~V. a nd~{Larionova}, L.~V., \&
  {Marscher}, A.~P. 2008, \apj, 672, 40

\bibitem[{{Ikejiri} {et~al.}(2011){Ikejiri}, {Uemura}, {Sasada}, {Ito},
  {Yamanaka}, {Sakimoto}, {Arai}, {Fukazawa}, {Ohsugi}, {Kawabata}, {Yoshida},
  {Sato}, \& {Kino}}]{2011PASJ...63..639I}
{Ikejiri}, Y., {et~al.} 2011, \pasj, 63, 639

\bibitem[{{Isobe} {et~al.}(2010){Isobe}, {Sugimori}, {Kawai}, {Ueda}, {Negoro},
  {Sugizaki}, {Matsuoka}, {Daikyuji}, {Eguchi}, {Hiroi}, {Ishikawa},
  {Ishiwata}, {Kawasaki}, {Kimura}, {Kohama}, {Mihara}, {Miyoshi}, {Morii},
  {Nakagawa}, {Nakahira}, {Nakajima}, {Ozawa}, {Sootome}, {Suzuki}, {Tomida},
  {Tsunemi}, {Ueno}, {Yamamoto}, {Yamaoka}, {Yoshida}, \& {MAXI
  Team}}]{2010PASJ...62L..55I}
{Isobe}, N., {et~al.} 2010, \pasj, 62, L55

\bibitem[{{Itoh} {et~al.}(2013){Itoh}, {Tanaka}, {Fukazawa}, {Kawabata},
  {Kawaguchi}, {Moritani}, {Takaki}, {Ueno}, {Uemura}, {Akitaya}, {Yoshida},
  {Ohsugi}, {Hanayama}, {Miyaji}, \& {Kawai}}]{2013ApJ...775L..26I}
{Itoh}, R., {et~al.} 2013, \apjl, 775, L26

\bibitem[{{Kawabata} {et~al.}(2008){Kawabata}, {Nagae}, {Chiyonobu}, {Tanaka},
  {Nakaya}, {Suzuki}, {Kamata}, {Miyazaki}, {Hiragi}, {Miyamoto}, {Yamanaka},
  {Arai}, {Yamashita}, {Uemura}, {Ohsugi}, {Isogai}, {Ishitobi}, \&
  {Sato}}]{2008SPIE.7014E.151K}
{Kawabata}, K.~S., {et~al.} 2008, in Society of Photo-Optical Instrumentation
  Engineers (SPIE) Conference Series, Vol. 7014, Society of Photo-Optical
  Instrumentation Engineers (SPIE) Conference Series

\bibitem[{{Lico} {et~al.}(2014){Lico}, {Giroletti}, {Orienti}, {G{\'o}mez},
  {Casadio}, {D'Ammando}, {Blasi}, {Cotton}, {Edwards}, {Fuhrmann}, {Jorstad},
  {Kino}, {Kovalev}, {Krichbaum}, {Marscher}, {Paneque}, {Piner}, \&
  {Sokolovsky}}]{2014A&A...571A..54L}
{Lico}, R., {et~al.} 2014, \aap, 571, A54

\bibitem[{{Lockman} \& {Savage}(1995)}]{1995ApJS...97....1L}
{Lockman}, F.~J., \& {Savage}, B.~D. 1995, \apjs, 97, 1

\bibitem[{{Marscher}(2014)}]{2014ApJ...780...87M}
{Marscher}, A.~P. 2014, \apj, 780, 87

\bibitem[{{Marscher} {et~al.}(2008){Marscher}, {Jorstad}, {D'Arcangelo},
  {Smith}, {Williams}, {Larionov}, {Oh}, {Olmstead}, {Aller}, {Aller},
  {McHardy}, {L{\"a}hteenm{\"a}ki}, {Tornikoski}, {Valtaoja}, {Hagen-Thorn},
  {Kopatskaya}, {Gear}, {Tosti}, {Kurtanidze}, {Nikolashvili}, {Sigua},
  {Miller}, \& {Ryle}}]{2008Natur.452..966M}
{Marscher}, A.~P., {et~al.} 2008, \nat, 452, 966

\bibitem[{{Massaro} {et~al.}(2004){Massaro}, {Perri}, {Giommi}, \&
  {Nesci}}]{2004A&A...413..489M}
{Massaro}, E., {Perri}, M., {Giommi}, P., \& {Nesci}, R. 2004, \aap, 413, 489

\bibitem[{{Mead} {et~al.}(1990){Mead}, {Ballard}, {Brand}, {Hough}, {Brindle},
  \& {Bailey}}]{1990A&AS...83..183M}
{Mead}, A.~R.~G., {Ballard}, K.~R., {Brand}, P.~W.~J.~L., {Hough}, J.~H.,
  {Brindle}, C., \& {Bailey}, J.~A. 1990, \aaps, 83, 183

\bibitem[{{Moore} {et~al.}(1982){Moore}, {Angel}, {Duerr}, {Lebofsky},
  {Wisniewski}, {Rieke}, {Axon}, {Bailey}, {Hough}, \&
  {McGraw}}]{1982ApJ...260..415M}
{Moore}, R.~L., {et~al.} 1982, \apj, 260, 415

\bibitem[{{Morozova} {et~al.}(2014){Morozova}, {Larionov}, {Troitsky},
  {Jorstad}, {Marscher}, {G{\'o}mez}, {Blinov}, {Efimova}, {Hagen-Thorn},
  {Hagen-Thorn}, {Joshi}, {Konstantinova}, {Kopatskaya}, {Larionova},
  {Larionova}, {L{\"a}hteenm{\"a}ki}, {Tammi}, {Rastorgueva-Foi}, {McHardy},
  {Tornikoski}, {Agudo}, {Casadio}, {Molina}, {Volvach}, \&
  {Volvach}}]{2014AJ....148...42M}
{Morozova}, D.~A., {et~al.} 2014, \aj, 148, 42

\bibitem[{{Nolan} {et~al.}(2012){Nolan}, {Abdo}, {Ackermann}, {Ajello},
  {Allafort}, {Antolini}, {Atwood}, {Axelsson}, {Baldini}, {Ballet}, \&
  et~al.}]{2012ApJS..199...31N}
{Nolan}, P.~L., {et~al.} 2012, \apjs, 199, 31

\bibitem[{{Piner} \& {Edwards}(2005)}]{2005ApJ...622..168P}
{Piner}, B.~G., \& {Edwards}, P.~G. 2005, \apj, 622, 168

\bibitem[{{Poole} {et~al.}(2008){Poole}, {Breeveld}, {Page}, {Landsman},
  {Holland}, {Roming}, {Kuin}, {Brown}, {Gronwall}, {Hunsberger}, {Koch},
  {Mason}, {Schady}, {vanden Berk}, {Blustin}, {Boyd}, {Broos}, {Carter},
  {Chester}, {Cucchiara}, {Hancock}, {Huckle}, {Immler}, {Ivanushkina},
  {Kennedy}, {Marshall}, {Morgan}, {Pandey}, {de Pasquale}, {Smith}, \&
  {Still}}]{2008MNRAS.383..627P}
{Poole}, T.~S., {et~al.} 2008, \mnras, 383, 627

\bibitem[{{Punch} {et~al.}(1992){Punch}, {Akerlof}, {Cawley}, {Chantell},
  {Fegan}, {Fennell}, {Gaidos}, {Hagan}, {Hillas}, {Jiang}, {Kerrick}, {Lamb},
  {Lawrence}, {Lewis}, {Meyer}, {Mohanty}, {O'Flaherty}, {Reynolds}, {Rovero},
  {Schubnell}, {Sembroski}, {Weekes}, \& {Wilson}}]{1992Natur.358..477P}
{Punch}, M., {et~al.} 1992, \nat, 358, 477

\bibitem[{{Schlegel} {et~al.}(1998){Schlegel}, {Finkbeiner}, \&
  {Davis}}]{1998ApJ...500..525S}
{Schlegel}, D.~J., {Finkbeiner}, D.~P., \& {Davis}, M. 1998, \apj, 500, 525

\bibitem[{{Schulz} \& {Lenzen}(1983)}]{1983A&A...121..158S}
{Schulz}, A., \& {Lenzen}, R. 1983, \aap, 121, 158

\bibitem[{{Spada} {et~al.}(2001){Spada}, {Ghisellini}, {Lazzati}, \&
  {Celotti}}]{2001MNRAS.325.1559S}
{Spada}, M., {Ghisellini}, G., {Lazzati}, D., \& {Celotti}, A. 2001, \mnras,
  325, 1559

\bibitem[{{Tramacere} {et~al.}(2009){Tramacere}, {Giommi}, {Perri},
  {Verrecchia}, \& {Tosti}}]{2009A&A...501..879T}
{Tramacere}, A., {Giommi}, P., {Perri}, M., {Verrecchia}, F., \& {Tosti}, G.
  2009, \aap, 501, 879

\bibitem[{{Tramacere} {et~al.}(2007){Tramacere}, {Massaro}, \&
  {Cavaliere}}]{2007A&A...466..521T}
{Tramacere}, A., {Massaro}, F., \& {Cavaliere}, A. 2007, \aap, 466, 521

\bibitem[{{Uemura} {et~al.}(2010){Uemura}, {Kawabata}, {Sasada}, {Ikejiri},
  {Sakimoto}, {Itoh}, {Yamanaka}, {Ohsugi}, {Sato}, \&
  {Kino}}]{2010PASJ...62...69U}
{Uemura}, M., {et~al.} 2010, \pasj, 62, 69

\bibitem[{{Urry} \& {Padovani}(1995)}]{1995PASP..107..803U}
{Urry}, C.~M., \& {Padovani}, P. 1995, \pasp, 107, 803

\bibitem[{{Urry} {et~al.}(2000){Urry}, {Scarpa}, {O'Dowd}, {Falomo}, {Pesce},
  \& {Treves}}]{2000ApJ...532..816U}
{Urry}, C.~M., {Scarpa}, R., {O'Dowd}, M., {Falomo}, R., {Pesce}, J.~E., \&
  {Treves}, A. 2000, \apj, 532, 816

\bibitem[{{Ushio} {et~al.}(2010){Ushio}, {Stawarz}, {Takahashi}, {Paneque},
  {Madejski}, {Hayashida}, {Kataoka}, {Tanaka}, {Tanaka}, \&
  {Ostrowski}}]{2010ApJ...724.1509U}
{Ushio}, M., {et~al.} 2010, \apj, 724, 1509

\bibitem[{{Villata} {et~al.}(1998){Villata}, {Raiteri}, {Lanteri}, {Sobrito},
  \& {Cavallone}}]{1998A&AS..130..305V}
{Villata}, M., {Raiteri}, C.~M., {Lanteri}, L., {Sobrito}, G., \& {Cavallone},
  M. 1998, \aaps, 130, 305

\bibitem[{{Visvanathan} \& {Wills}(1998)}]{1998AJ....116.2119V}
{Visvanathan}, N., \& {Wills}, B.~J. 1998, \aj, 116, 2119

\bibitem[{{Wright}(2006)}]{2006PASP..118.1711W}
{Wright}, E.~L. 2006, \pasp, 118, 1711

\bibitem[{{Zhang} {et~al.}(2002){Zhang}, {Treves}, {Celotti}, {Chiappetti},
  {Fossati}, {Ghisellini}, {Maraschi}, {Pian}, {Tagliaferri}, \&
  {Tavecchio}}]{2002ApJ...572..762Z}
{Zhang}, Y.~H., {et~al.} 2002, \apj, 572, 762

\end{thebibliography}

\end{document}